\documentstyle[epsf]{mn}

\title[X-ray variability in NGC 2516]{A search for X-ray variability in the open cluster NGC 2516}

\author[Ramsay, Harra \& Kay]{
Gavin Ramsay, Louise Harra, Hilary Kay\\
Mullard Space Science Laboratory, University College London,
Holmbury St. Mary, Dorking, Surrey, RH5 6NT, UK\\}
\date{Received: }

\begin{document}
\outer\def\gtae {$\buildrel {\lower3pt\hbox{$>$}} \over 
{\lower2pt\hbox{$\sim$}} $}
\outer\def\ltae {$\buildrel {\lower3pt\hbox{$<$}} \over 
{\lower2pt\hbox{$\sim$}} $}
\newcommand{\ergscm} {ergs s$^{-1}$ cm$^{-2}$}
\newcommand{\ergss} {ergs s$^{-1}$}
\newcommand{\ergsd} {ergs s$^{-1}$ $d^{2}_{100}$}
\newcommand{\pcmsq} {cm$^{-2}$}
\newcommand{\ros} {\sl ROSAT}
\newcommand{\exo} {\sl EXOSAT}
\newcommand{\chandra} {\sl Chandra}
\newcommand{\xmm} {\sl XMM-Newton}
\def\rchi{{${\chi}_{\nu}^{2}$}}
\newcommand{\Msun} {$M_{\odot}$}
\newcommand{\Mwd} {$M_{wd}$}
\def\Mdot{\hbox{$\dot M$}}
\def\mdot{\hbox{$\dot m$}}

\maketitle

\begin{abstract}

We present the results of a search for X-ray variability in the
galactic open cluster NGC 2516. This cluster has been observed on 8
occasions using {\xmm} making it an excellent target for such a
study. We find 4 sources which show significant variability, implying
the detection of one significantly variable source every 25ksec. All
sources exhibit spectra which can be fitted using an absorbed one or
two temperature thermal plasma model. The brightest of these flares
also show a prominent Iron line near $\sim$7.0keV. All 4 sources lie
near the observed main sequence of NGC 2516 suggesting they are
cluster members. We propose that at least 3 of the 4 objects are RS
CVn systems. We compare the properties of the brightest flare with
those of solar flares.

\end{abstract}


\section{Introduction}

NGC 2516 is a rich, well studied, galactic open cluster. It has a low
metallicity (eg Cameron 1985, Jeffries, James \& Thurston 1998), is
nearby ($\sim$387pc; Jeffries, Thurston \& Pye 1997), is young
($\sim10^{8}$ years; eg Feinstein, Marraco, Mirabel et al 1973) and
has a relatively low line of sight extinction. Because of its low
metallicity, it is an excellent target to test how coronal activity
and stellar structure depend on metal content.

NGC 2516 has been the subject of several studies in the X-ray band. It
was observed using {\ros} by Dachs \& Hummel (1996), Jeffries,
Thurston \& Pye (1997) and Micela et al (2000). More recently it was
the subject of a study using {\xmm} (Sciortino et al 2001) and also
{\chandra} (Harnden et al 2001). These studies have concentrated on
determining the cluster X-ray luminosity function: Sciortino et al
(2001) found that dwarf G and K stars in NGC 2516 were less luminous
than those in the Pleiades (which has a similar age but solar
metallicity ratio). In contrast, the luminosity function of dwarf M
stars in NGC 2516 and the Pleiades were indistinguishable.

In this paper, we show the results of a search for X-ray variable
objects in NGC 2516 using {\xmm} data. With the relatively large
number of X-ray objects it is an excellent target to determine the
frequency of stellar flaring activity in a young, metal poor
cluster. Further, NGC 2516 has been observed on 8 occasions with
{\xmm} since it has been used as a calibration target.

\begin{table}
\begin{center}
\begin{tabular}{lrcr}
\hline
Date & Orbit & ObsId & Clean Time\\
     &     &       &  (ksec) \\
\hline
2000-04-06 & 060 & 0113891001 & 17.6\\
2000-04-06 & 060 & 0113891101 & 15.0\\
2000-06-10 & 092 & 0126511201 & 26.1\\
2000-10-31 & 164 & 0113891201 & 0.5\\
2001-01-29 & 209 & 0134531201 & 17.0\\
2001-01-29 & 209 & 0134531301 & 5.0\\
2001-06-24 & 272 & 0134531301 & 0.0\\
2001-10-29 & 346 & 0134531501 & 16.7\\
\hline
\end{tabular}
\end{center}
\caption{The observation log of {\xmm} observations of NGC 2516. We
show the {\xmm} orbit number and the observation id.}
\label{obs}
\end{table}

\section{Observations}

The satellite {\xmm} was launched in Dec 1999 by the European Space
Agency. It has the largest effective area of any imaging X-ray
satellite. The EPIC instruments have imaging detectors covering the
energy range 0.15--10keV with moderate spectral resolution. The
observation log is shown in Table \ref{obs}. The first two
observations were centered at 7$^h$ 58$^m$ 20$^s$, --60$^{\circ}$
52$^{'}$ 13$^{''}$ (2000), and the rest at 7$^h$ 58$^m$ 22$^s$,
-60$^{\circ}$ 45$^{'}$ 36$^{''}$ (2000).  (We do not discuss the data
taken using the Reflection Grating Spectrometer since the sources were
comparatively faint and source confusion would be a problem. The
Optical Monitor was in blocked mode).

\begin{figure*}
\begin{center}
\setlength{\unitlength}{1cm}
\begin{picture}(8,7.5)
\put(-5.,-3.5){\includegraphics{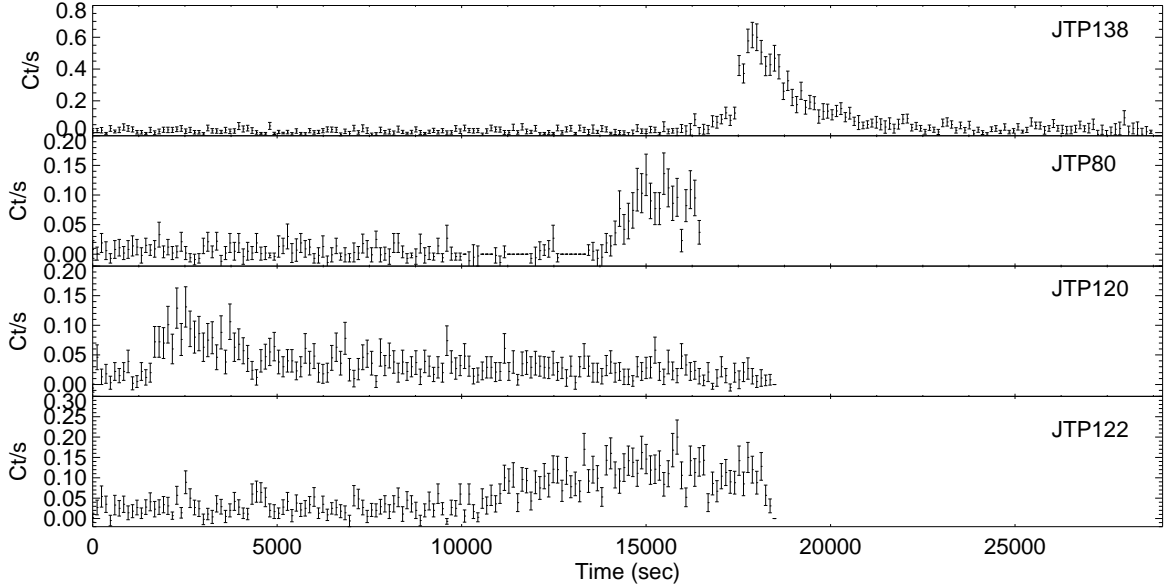}}
\end{picture}
\end{center}
\caption{The background subtracted X-ray light curves for each of the
4 sources which showed significant variability. The bin size is 120
sec and covers the energy range 0.2-10.0keV.}
\label{light} 
\end{figure*}

The data were processed using the {\sl XMM-Newton} {\sl Science
Analysis Software} (SAS) v5.3.3. An image was generated from each
dataset in the energy range 0.2-5.0keV. We removed data at the edge of
the detector furthest from on-axis since these rows had increased
noise. We also selected those events which were tagged as {\tt
FLAG=0}: this excludes events close to, for instance, hot pixels and
chip edges. Time intervals which showed high levels of particle
background were excluded. The resulting images were then source
searched using standard SAS tasks. Events were then extracted around a
specified radius from the source positions. The events for each source
were then subject to a Discrete Fourier Transform in the time interval
of 1min to 0.7$\times$Exposure length. All power spectra which showed
a peak above a certain criteria above the mean power were flagged and
inspected by eye. We also had a second flagging procedure which
indicated which objects had a count rate which varied over a certain
value.

After the completion of this process, 4 sources were found to show
significant variability in their X-ray light curves. Their X-ray
positions were compared to the list of X-ray sources in NGC 2516 and
their corresponding optical counterparts (Jeffries, Thurston \& Pye
1997). We found counterparts for all 4 sources: the offset between the
{\xmm} and optical counterpart was typically better than 6$^{''}$
(which is the FWHM of sources in the EPIC pn). We show their X-ray
positions along with their counterpart of the X-ray catalogue of
Jeffries et al (1997) in Table \ref{sources}. For each source we
examined the other datasets to determine if the source was also
detected at other epochs. In each case the source was detected but at
low count rates (typically 0.01-0.02 ct/s). However, they showed no
significant variation in these other datasets.

\begin{figure}
\begin{center}
\setlength{\unitlength}{1cm}
\begin{picture}(8,10)
\put(-1,-3){\includegraphics{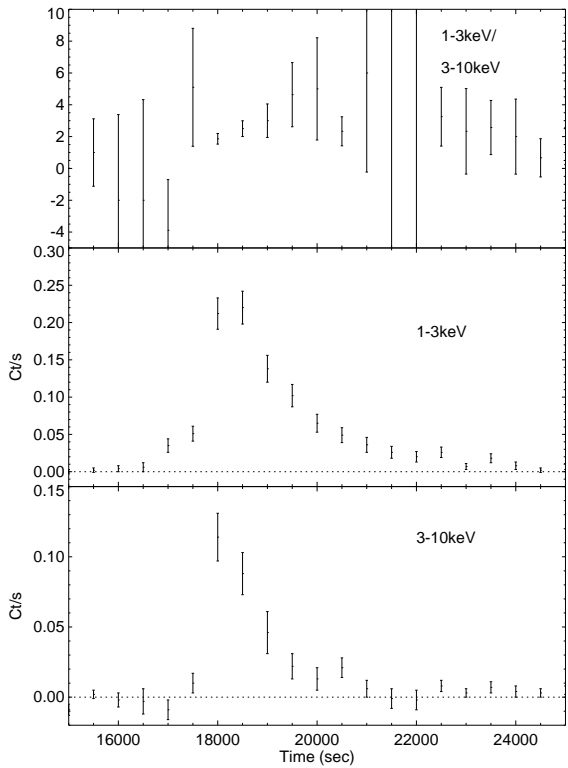}}
\end{picture}
\end{center}
\caption{The flare seen in JTP138 in the 1--3keV (middle panel) and
3--10keV (bottom panel) energy band and the 1--3keV/3--10keV softness
ratio (top panel).}
\label{hard} 
\end{figure}

We show the (background subtracted) X-ray light curves for each source
in Figure \ref{light}. Three of the sources show evidence for a rapid
flare, although in one of those (JTP80) the observation ends shortly
after the start of the flare. The other source (JTP122) shows a more
gradual rise in brightness. The most dramatic source is JTP138 which
shows a gradual rise in flux lasting $\sim$12 min followed by a sharp
increase in flux taking place in less than 2 min. After reaching
maximum, it reaches half its maximum intensity after $\sim$20
mins. The duration of the flare lasts around 90 min. We show in Figure
\ref{hard} the light curves in the 1--3keV and 3--10keV energy bands
for JTP 138 in time bins of 500 sec together with its softness ratio.
Although the hardness ratio curve shows some evidence for a softening
over the duration of the flare, it has large errors. However, there is
some evidence for a small increase in flux in the 1--3keV light curve
just before the sharp rise that is not seen in the 3--10keV light
curve, although again this may just be due to the low signal to noise.

\begin{table}
\begin{tabular}{rccrcc}
\hline
JTP & RA & Dec & $V$ & $(B-V)$ & ObsId\\
\hline
80 & 07 57 58.5 & -60 56 53 & 12.84 & 0.63 & 0113891101\\
138 & 07 59 20.1 & -60 34 44& 11.96 & 0.51 & 0126511201\\
114 & 07 58 43.3 & -60 55 24 & 13.97 & 0.87 & 0134531501\\
122 & 07 58 50.8 & -60 38 35 & 9.51 & 0.08 & 0134531501\\
\hline
\end{tabular}
\caption{The X-ray sources which showed significant X-ray
variability. JTP refers to the X-ray source number of Jeffries,
Thurston \& Pye (1997), the X-ray position (2000) is the position from
the XMM-Newton EPIC images and the optical counterparts are taken from
Jeffries et al. We also indicate the observation number (cf Table 1)
in which the source variation was detected.}
\label{sources}
\end{table}

\section{Source Spectra}

We extracted spectra from each of the 4 variable sources: only the
time interval of the flare duration and single and double events were
used. We also extracted background spectra from a nearby source free
area off the detector. We used the `canned' response files appropriate
to the location of the source on the detector and generated an
auxiliary file using the SAS task {\tt arfgen}. We fitted the spectra
using the X-ray fitting package {\tt XSPEC} (Dorman \& Arnaud 2001).

\begin{figure}
\begin{center}
\setlength{\unitlength}{1cm}
\begin{picture}(8,6)
\put(-1.2,-0.8){\includegraphics{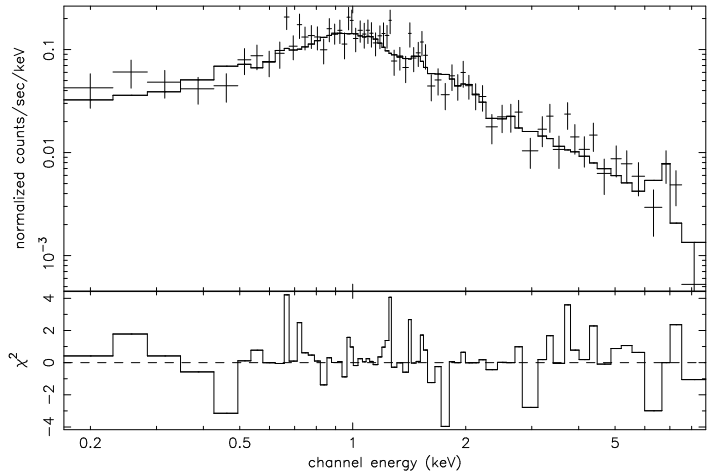}}
\end{picture}
\end{center}
\caption{The EPIC pn spectra of source JTP138. The best fit - shown as
a solid line - is an absorbed two temperature plasma model.}
\label{spec} 
\end{figure}

All spectra could be fitted using an absorbed one or two temperature
plasma model (we used the {\tt MEKAL} model in {\tt XSPEC}). We show
the spectrum of the brightest flare (JTP138) in Figure \ref{spec}: the
fit is good with no evidence for significant residuals at particular
energies. This source also showed evidence for an Fe line at
7.0$\pm0.2$keV (the other sources did not). We show the spectral fits
in Table \ref{fits}.

All spectra show a moderate amount of absorption, typically
$\sim5\times10^{20}$ \pcmsq, and temperatures of a few keV.  In
determining their luminosity we assume that all 4 sources are members
of NGC 2516 and its distance is 387 pc (Jeffries, Thurston \& Pye
1997). In the case of JTP138, where we observe the total duration of
the flare, the energy released was $8\times10^{34}$ ergs. In the case
of JTP80, where we observed only the start of the flare, the energy
released was greater than $1\times10^{34}$ ergs. There was a similar
energy release observed in JTP114 where most of the flare was
observed.

\begin{table*}
\begin{tabular}{rccrrrrr}
\hline
JTP & $N_{H}$ & $kT_{1}$ & $kT_{2}$ & Observed flux & \multicolumn{2}{c}
{Bolometric luminosity} & \rchi\\
    & $\times10^{20}$ \pcmsq & (keV) & (keV) & \ergscm & 10$^{30}$
\ergss & 10$^{34}$ ergs & (dof)\\
\hline
80 &  5.4$^{+6.0}_{-4.3}$& 3.1$^{+2.1}_{-0.6}$ & &
3.1$^{+0.5}_{-0.6}\times10^{-13}$ & 5.5$^{+0.9}_{-1.1}$ & 
1.2$\pm{0.2}$ & 0.64 (39)\\
138 & 9.5$^{+5.0}_{-3.5}$ & 0.8$^{+0.4}_{-0.5}$ & 5.3$^{+3.1}_{-1.4}$ & 
1.05$\pm0.14\times10^{-13}$ & 1.9$^{+2.1}_{-1.6}$ &
8.0$^{+0.8}_{-1.3}$ & 0.97 (64)\\
114 & 7$^{+8}_{-6}$ & 1.8$^{+0.9}_{-0.3}$& & 1.64$\times10^{-13}$ &
4.9$^{+0.8}_{-1.3}$ & 1.1$^{+0.2}_{-0.3}$ & 1.27 (16)\\
122 & 5.5$^{+15}_{-4.5}$ & 1.0$^{+0.7}_{-0.5}$ &
3.7$^{+2.7}_{-1.7}$ & 3.0$\pm0.6\times10^{-13}$ & 8.0$^{+1.2}_{-1.4}$ &
5.6$^{+0.8}_{-1.0}$ & 1.67 (33)\\
\hline
\end{tabular}
\caption{The spectral fits to the four X-ray varying sources. We have
extracted events covering only flare time interval. When determining
the luminosity of the flare we assume a distance of 387pc and a solar
metallicity.}
\label{fits}
\end{table*}

\section{The nature of the sources}

\begin{figure}
\begin{center}
\setlength{\unitlength}{1cm}
\begin{picture}(8,5)
\put(-1.2,-0.5){\includegraphics{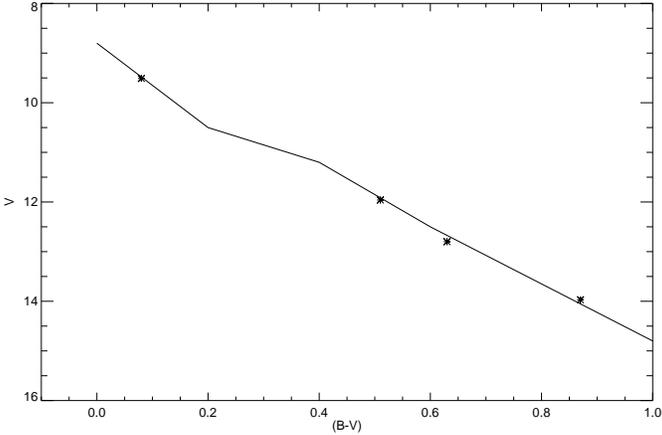}}
\end{picture}
\end{center}
\caption{We plot the location of the optical counterparts of the X-ray
flaring sources in the $V,B-V$ plane. We also show the location of the
main sequence in NGC 2516 taken from the data of Sung et al (2002).}
\label{main} 
\end{figure}

We have identified each of the 4 X-ray variable sources with an
optical counterpart (Table \ref{sources}). Jeffries et al (1997)
identify JTP122 as having an AOV spectral type: the other 3 sources
have not been classified spectroscopically. Further, JTP138 and JTP122
were both classed as cluster members by King (1978) from a proper
motion survey. To determine the likelihood of the other two sources
being cluster members we plotted the $V, B-V$, data of Sung et al
(2002) and traced the position of the main sequence. We then
over-plotted the optical colours of our X-ray variable sources (Figure
\ref{main}). All 4 sources lie close to the main sequence of NGC 2516:
we conclude that each source is likely to be a cluster member.

Stellar coronal activity is common in stars which have a convective
layer and a relatively rapid rotation period. Enhanced levels of
activity are often seen in close binary systems, for example the RS CVn
systems which typically consist of a giant or sub-giant primary with a
main-sequence or sub-giant companion. In many RS CVn systems the
rotation of one of the components is synchronised with the orbital
motion, leading to enhanced rotation and high levels of activity, seen
as large spot coverage and flares in X-ray, optical and radio
bands. Additionally X-ray activity has also been seen on early-type
stars, such as Castor A and B. Here it is thought the flaring
component may originate from low-mass, late-type companions (G\"{u}del
at al. 2001). X-ray flares have also been reported in pre-main
sequence stars (eg Schmidt 1994) and single solar type stars (eg Landini
et al 1986).

The energy released in the X-ray flares we observed in NGC 2516 were
$\sim\times10^{34-35}$ ergs and lasted around 1--2 hrs. This contrasts
to the energy released on a flare seen on Algol (7$\times10^{36}$
ergs) which lasted over a day (Ottmann \& Schmitt 1996).  In dwarf M
stars flares have been seen in the $U$ band (eg Byrne, Doyle \& Butler
1984) although they are relatively rare and last several 10's of
minutes.  In X-rays, dwarf M stars show flares of energies typically
$\sim10^{30-34}$ ergs with a duration less than 1 hour (eg Cheng \&
Pallavicini 1991). We consider that it is unlikely that the flares
originate from pre-main sequence stars since NGC 2516 is not a star
forming region. Solar type stars have also been found to show flares
in various energy bands with energies $\sim10^{33-38}$ ergs, although
they are very rare (eg Schaefer et al 2000).

In contrast, flares have been seen on RS CVn systems which share
similar characteristics to the flares we have observed. van den Oord,
Mewe \& Brinkman (1988) estimated the energy released in a flare on
the RS CVn system $\sigma^{2}$ CrB was 2.4$\times10^{34}$ ergs and
lasted $\sim$ 90 mins. Further, they found clear evidence for a
softening of the light curve after the sharp rise in flux. This is
comparable to the RS CVn system II Peg which released
$>5\times10^{34}$ ergs in a flare lasting more than several hours
(Doyle et al 1991, Doyle, van der Oord \& Kellett 1992). Their X-ray
spectra are well modelled using a thermal plasma of a few
keV. Further, there are several instances where Fe K$\alpha$ emission
has been seen in stars with active stellar coronae (eg Tsuru et al
1989, G\"{u}del et al 2001).

We suggest that JTP 138, JTP 80 and JTP 114 are good candidates for RS
CVn systems. The fact that JTP 122 has been classified as AO V
spectral type may indicate that it has a dwarf M star as a companion.
Followup spectroscopy will be able to test these assertions.

\section{The frequency of flaring activity}

NGC 2516 has been observed using {\xmm} for 97.4 ksec (cf Table 1).
We detected 4 significantly variable X-ray sources over this time
implying the detection of one X-ray variable source every
$\sim$25ksec. The first two observations were offset from the other
observations by $\sim7^{'}$ (cf Table 1). In this first field we
detected 113 X-ray sources and in the second 151 (using the longest
exposures in these fields). We detected 1 variable source in field 1
and 3 in field 2. This gives 2.7$\times10^{-4}$ variable X-ray sources
per ksec per X-ray source in field 1 and 3.0$\times10^{-4}$ in field
2. Although simplistic, if every source in the field was a potential
X-ray variable source, then these rates would imply that each source
would show a flare every $\sim$40 days.

\section{Comparison with solar flares}

Flaring activity has been seen on the Sun at many wavelengths and
timescales. We have searched for solar flares that have a similar
temporal profile to `JTP 138' using the Geostationary Operational
Environmental Satellites (GOES). The satellites are in geostationary
orbit, their purpose being to monitor the Sun's X-ray emission
continuously.

\begin{figure}
\begin{center}
\setlength{\unitlength}{1cm}
\begin{picture}(8,4.5)
\put(-1,-1.){\includegraphics{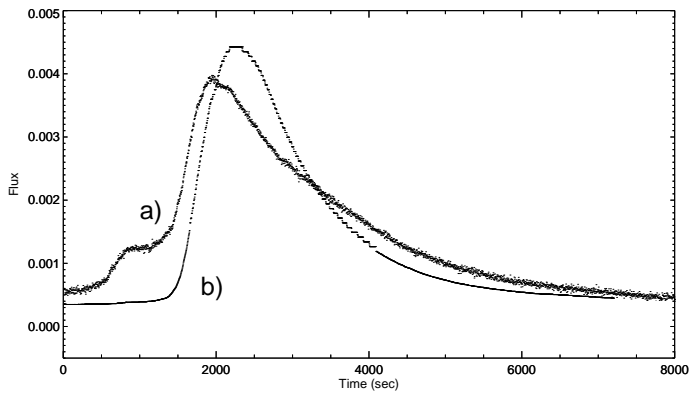}}
\end{picture}
\end{center}
\caption{Two solar flares in the wavelength range (1-8\AA). Flare a)
which took place on 3 Sep 1996 is a relatively small solar flare,
which lasts for approximately 1.5 hrs. It shows prominent pre-flare
activity. The flux has been multiplied by 10. Flare b) (12 Feb 2001)
is an average intensity solar flare. The units for flux is
ergs/s/cm$^{-2}$.}
\label{solar} 
\end{figure}

\begin{figure}
\begin{center}
\setlength{\unitlength}{1cm}
\begin{picture}(8,5.5)
\put(-1.5,-7.){\includegraphics{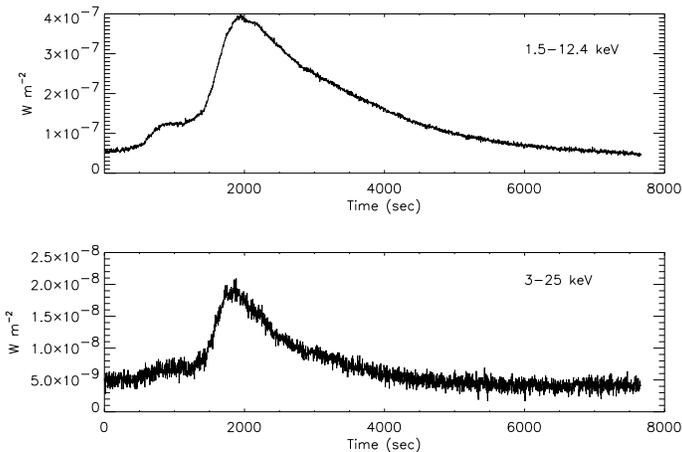}}
\end{picture}
\end{center}
\caption{We show the light curve of the flare a in Figure \ref{solar}
split into soft and hard X-ray bands. Like the flare seen in JTP 138
the hard flux drops more rapidly than at softer energies and maximum
hard flux is reached earlier than at softer energies.}
\label{solar2} 
\end{figure}

We have identified several solar flares which have similar timescales
to that seen in JTP 138. Figure \ref{solar} (curve a) shows the light
curve of a `small' solar flare which has an intensity several orders
of magnitude lower than the largest flares in the current solar
cycle. Like JTP 138 it shows a small increase in flux (lasting
$\sim$10 mins) before a rapid rise which takes around 10 mins to reach
maximum flux. We have estimated its temperature using ratios of the
low energy channel on GOES (1-8\AA, 1.5--12.4keV) with the high energy
channel (0.5-4\AA, 3--25keV). The temperature reached a maximum of 0.6
keV, with a luminosity of 7$\times$10$^{25}$ \ergss. The second
example solar flare is also shown in Figure \ref{solar} (curve
b). This flare is brighter, and reaches a temperature of 0.8 keV, with
a luminosity of 2$\times$10$^{26}$ \ergss. In contrast to the small
solar flare and JTP 138, there is no evidence for a pre-flare increase
in flux. Both solar flares last around 1--1.5 hrs, similar to JTP
138. In Figure \ref{solar2} we show the energy resolved light curve of
the small solar flare. Similar to the flare seen in JTP 138, the solar
flare reaches a maximum in hard X-rays earlier than in soft X-rays.

Although the shape and duration of these solar flares are similar to
that of JTP 138, they have lower temperatures and much lower
luminosities. The luminosity may simply be a reflection of the volume
of the emitting plasma. In the case of RS CVn systems, the flare size
can extend to the size of the binary system (or indeed, originate in
the intra-binary region as seen in the RS CVn system CF Tuc, Gunn et
al 1997). For reasonable system parameters and an orbital period of 2
days, the separation is $\sim5\times10^{11}$ cm.  Moreover, the Suns
rotation rate has been slowing down since it joined the main
sequence. It therefore has lower magnetic activity compared to stars
which have recently joined the main sequence (as in the case of stars
in NGC 2516).

Numerous explanations for the triggering mechanism of solar flares
have been developed over the years. The most popular current theory
for converting magnetic energy into kinetic and thermal energies is
that of magnetic reconnection (eg Priest \& Forbes 1999). The site of
the reconnection takes place high in the solar corona above magnetic
loops, or between at least two loop systems that interact. At the
reconnection site, the electrons are accelerated and move along the
loops with a high velocity ($v\sim0.5c$). Some of these electrons will
bombard the chromosphere which is denser and cause this plasma to be
heated. This plasma will then evaporate into the loops causing the
soft X-ray emission such as is shown in Figure 5 (cf Harra 2000 for
more details). Although the details are still uncertain, observations
support this view.

Feldman, Laming \& Doschek (1995) found a correlation between solar
and stellar flare temperatures and emission measures. Shibata \&
Yokoyama (1999) proposed a model based on a magnetic reconnection
model with heat conduction and chromospheric evaporation to account
for this correlation. Based on our spectral fits to JTP 138, we find
an emission measure of 7$\times10^{53}$ cm$^{-3}$ for JTP 138. For a
temperature of 6$\times10^{7}$ K for the hot component, Figure 1 of
Shibata \& Yokoyama (1999) suggests a magnetic field strength of
$\sim$100 G for this source. Measuring stellar magnetic field
strengths is difficult. However, observations of various RS CVn
systems show evidence for magnetic field strengths of the order
$\sim$100--1000 G (eg Donati 1999), indicating that a field strength
of 100 G for JTP 138 is reasonable.

\section{Conclusions}

We have performed a X-ray time variability survey of the open cluster
NGC 2516. We have found 4 systems which have shown significant
variability. Three of these systems show light curves which are
characteristic of stellar flaring activity. Based on the flare
duration and their spectral properties we suggest that they are RS CVn
systems. The fourth system has been identified with a AO V type star
and propose that it may have an unseen dwarf M star which shows
flaring activity. An optical spectroscopic study of these objects is
required. Finally, this study shows that open clusters are excellent
objects to search for stellar flaring activity and may be relatively
common.

\section{Acknowledgments}

This paper is Based on observations obtained with XMM-Newton, an ESA
science mission with instruments and contributions directly funded by
ESA Member States and the USA (NASA). We thank Mat Page for the use of
some of his software. We acknowledge the use of the WEBDA database of
galactic clusters.

\end{document}